\documentstyle[12pt]{article}
\newcommand{\be}{\begin{equation}}
\newcommand{\ee}{\end{equation}}
\newcommand{\bea}{\begin{eqnarray}}
\newcommand{\ena}{\end{eqnarray}}
\newcommand{\vs}[1]{\rule[- #1 mm]{0mm}{#1 mm}}

\newcommand{\MSB}{\overline {MS}}

\newcommand{\alfapi}{{\alpha \over 2 \pi}}

\newcommand{\alfspi}{ {\alpha_s(\mu) \over 2 \pi} }
\newcommand{\alfsmpi}{ {\alpha_s(M) \over 2 \pi} }
\newcommand{\alfsqpi}{ {\alpha_s(Q) \over 2 \pi} }
\newcommand{\alfspis}{ \left( {\alpha_s(\mu) \over 2 \pi} \right) ^2 }

\newcommand{\dsigd}{  {d \sigma^{D} \over {d {\vec p}_T d \eta}}}
\newcommand{\dsigsf}{{d \sigma^{SF} \over {d {\vec p}_T d \eta}}}

\newcommand{\dsigig}{{d \sigma^{i \gamma \rightarrow jet}
            \over {d {\vec p}_T d \eta}}}

\newcommand{\dsigij}{{d \sigma^{i j \rightarrow jet}
            \over {d {\vec p}_T d \eta}}}

\newcommand{\NP}[1]{Nucl.\ Phys.\ {\bf #1}}
\newcommand{\PL}[1]{Phys.\ Lett.\ {\bf #1}}

\newcommand{\PR}[1]{Phys.\ Rev.\ {\bf #1}}
\newcommand{\PRL}[1]{Phys.\ Rev.\ Lett.\ {\bf #1}}
\newcommand{\ZPH}[1]{Z.\ Phys.\  {\bf #1}}

\begin{document}

\renewcommand{\thefootnote}{\fnsymbol{footnote}}
\newpage
\setcounter{page}{0}

\vs{40}

\begin{center}
{\Large {\bf{On the photoproduction of jets at HERA}}} \\
\vspace{0.7 cm}
{\large P. Aurenche, J.-Ph. Guillet} \\
{\em Laboratoire de Physique Th\'eorique ENS{\large{\em L}}APP
\footnote{URA 14-36 du
CNRS, associ\'ee \`a l'Ecole Normale Sup\'erieure de Lyon, et au
Laboratoire d'Annecy-le-Vieux de Physique des Particules.} $-$ Groupe
d'Annecy\\
LAPP, IN2P3-CNRS, B.P. 110, F-74941 Annecy-le-Vieux Cedex, France}
\\[0.7cm]
{\large M. Fontannaz} \\
{\em Laboratoire de Physique Th\'eorique et Hautes Energies
\footnote {Laboratoire associ\'e au CNRS (URA 63). } \\
Universit\'e de Paris XI, b\^atiment 211, F-91405 Orsay Cedex, France}
\\[0.7cm]
\end{center}
\vs{20}

\centerline{ \bf{Abstract}}
\vs{3}

We discuss the inclusive jet production at HERA in the next-to-leading
logarithm approximation. Theoretical uncertainties are considered in
some  details. We show the importance of the jet rapidity distribution
to constrain  the parton densities in the photon. A comparison is made
with the recent H1  data.

\vs{15}

\rightline{ENSLAPP-A-484/94}
\rightline{LPTHE Orsay 94-80}
\rightline{August 1994}


\newpage

\indent

Several works have recently been devoted to the study of hard processes
in photon-proton scattering. The renewed interest in such reactions is
due to the opening up of a new kinematical domain at HERA and the
publication, for the first time, of data on large transverse momentum
jets in photoproduction  reactions at very high energies
\cite{h1jet}-\cite{h1part}.  The importance of this process arises from
the  possibility of probing the hadronic structure of the photon in a
way complementary to the usual studies of the deep inelastic photon
structure function. In the latter case, one can reach the very small $x$
domain of the parton densities in the photon and, in particular,
directly measure the quark  distributions whereas the gluon distribution
is constrained by the evolution equations. On the contrary, in
photoproduction reactions one directly probes the gluon density in the
photon assuming the parton distributions in the proton are known from
elsewhere. A quantitative study of both processes together should then
lead to a precise determination of the photon structure very much in the
same way as it does in purely hadronic reactions. As we shall see, the
non perturbative input to the photon structure function will also be
constrained by hard processes in photoproduction.

In the following we consider the next-to-leading logarithm predictions
of large $p_T$ jet production with emphasis on the interplay between
processes where the photon couples directly and those where it interacts
through its parton contents. Some theoretical uncertainties are
analysed. We then discuss in which kinematical domain the future HERA
data will be able to constrain the parton distributions. A comparison
with present data from H1 is performed and we discuss in particular the
role of the  gluon distribution. We stress the importance of the
rapidity distribution in  the determination of the photon stucture
functions. 

Several theoretical papers have already appeared on the subject
\cite{gorsto}, \cite{bks}. Contrary to previous works we stress the
phenomenological relevance of jet photoproduction data and indicate what
precision the data should reach and what rapidity domain should be
explored  to extract useful physics information.

\section{ The photon-proton cross section}

We first consider the scattering of a real photon off a proton at a
fixed center of mass energy $\sqrt s$. The cross section for the
production of a jet of transverse momentum $p_T$ and pseudo-rapidity
$\eta$ can be written  \be
\frac{d\sigma}{d\vec{p}_T d \eta} \ = \ \dsigd + \ \dsigsf 
\label{eq:sig}
\ee
where the first term on the right-hand side involves the "direct"
coupling of the photon to the hard partonic process while the second
term describes the interaction of the photon through its "structure
function". In the next-to-leading logarithmic approximation these terms
take the form  \bea
\dsigd (R) &=& \sum_{i=q,g} \int dx_{1} F_{i/p}(x_{1},M)    \nonumber \\
&\ & \ \ \ \ \ \alfspi \left( \dsigig + \ \alfspi K^{D}_{i \gamma} 
(R;M,\mu) \right)
\label{eq:dir}
\ena
and
\bea
\dsigsf (R) &=& \sum_{i,j=q,g} \int dx_{1} dx_{2}
 F_{i/p}(x_{1},M) F_{j/\gamma}(x_{2},M) \nonumber \\
&\ & \ \ \ \ \ \ \ \
\alfspis \left( \dsigij \ + \ \alfspi K^{SF}_{ij} (R;M,\mu) \right)
\label{eq:sf}
\ena
which exhibit the higher order corrections to the hard processes
$K^{D}_{i\gamma} (R;M)$ and $K^{SF}_{i j} (R;M,\mu)$ respectively
\cite{bogre}.  The latter term is the same as in purely hadronic
reactions and has been calculated some time ago in  \cite{guil} while
the former term has recently been computed and used  in photon-photon
collisions \cite{gaga}. For consistency, the parton distributions in the
proton, $ F_{i/p}(x,M) $, and in the photon,  $ F_{j/\gamma}(x,M) $,
also have to be defined beyond the leading-logarithm approximation.  The
parameter $R$ specifies the jet cone size, while $\mu$ and $M$ are  the
renormalization and factorization scales respectively.  The
factorization decomposition eqs. (\ref{eq:dir},\ref{eq:sf}),  involving
external photons, and has been derived in the $\MSB$ scheme
\cite{phopho} using the technics of \cite{cur} and \cite{rke}.

Since one of the aims of this work is the determination of the parton
distributions in the photon we now discuss the functions $
F_{i/\gamma}(x,M) $  in some details. They satisfy evolution equations
of type (we assume one flavor  for  simplicity)
\bea
{d F_{i/\gamma} \over d \ln M^2}&=& k_i + P_{ij} \otimes F_{j/\gamma},
\label{eq:evol}
\ena
with
\bea
P \otimes F_{i/\gamma}= \int^1_x dz\ P \left({x \over z} \right) 
\ F_{i/\gamma}(z)
\ena
Both the Altarelli-Parisi kernels $P_{ij}$ and the inhomogeneous terms
$k_i$ have a perturbative expansion in terms of the strong coupling
constant, \bea 
P_{ij} &=& \alfsmpi ( P^{(0)}_{ij} +  \alfsmpi  P^{(1)}_{ij} ) \nonumber \\
k_i &=& \alfapi ( k^{(0)}_i + \alfsmpi k^{(1)}_i )
\label{eq:ap}
\ena
The expressions for the various terms, in the $\MSB$ scheme, can be
found in  \cite{fupe} and in \cite{fopi}, \cite{glurevo} respectively.
The general solution of eqs. (\ref{eq:evol}) is written as a
superposition of two terms, \be
F_{i/\gamma}(x,M)  = F^{AN}_{i/\gamma}(x,M) + F^{NP}_{i/\gamma}(x,M)
\label{eq:solu}
\ee
where the first term on the right hand side satisfies the full
inhomogeneous equation and vanishes at some initial $M=M_{0}$ value
whereas the  "non-perturbative" term obeys the homogeneous evolution
equation. It will be discussed at length below.

The parton distributions are defined with respect to some   physical
process which is usually taken as the deep-inelastic photon structure
function $F^{\gamma}_2(x,Q)$. The relation between the structure
function and the parton distributions is given by: \bea
F^{\gamma}_2(x,Q) = F_{q/\gamma}(x,Q) ( 1+ \alfsqpi C^{(1)}_q(x)) +
F_{g/\gamma}(x,Q) \alfsqpi C^{(1)}_g(x) + C_{\gamma}(x)
\label{eq:strufu}
\ena
The Wilson coefficients $ C^{(1)}_i(x) $ have been calculated long time
ago \cite{barde}. We shall always assume in the following that they are
defined in the $\MSB$ scheme. The direct term $C_\gamma$ has also been
known for some time and in the $\MSB$ scheme it takes the form 
\bea
C_\gamma(x) = 6 \ \alfapi e^4_q \ \left( (x^2+(1-x)^2) \ln {1-x \over x} + 
8 x (1-x) -1 \right)
\label{eq:dirter}
\ena

When working beyond the leading logarithm approximation we are free to
choose the expression of $C_\gamma(x)$ by modifying the factorization
scheme. With the $\MSB$ choice, as in the above equation, we observe
that the $\ln(1-x)$ factor becomes very large near the boundary of phase
space and then it does not appear as a correction  in eq.
(\ref{eq:strufu}) since it may become numerically larger that the
leading terms $F_{i/\gamma}$ which are enhanced by a factor
$1/\alpha_s$. In order to keep the concept of a perturbative expansion
useful, it is proposed  in \cite{glurevo} to introduce the
$DIS_{\gamma}$ factorization scheme where the choice
\bea
C_\gamma(x)|_{DIS_\gamma} =0
\ena
is made. The parton distributions thus defined satisfy an evolution
equation of type (\ref{eq:evol}) with the inhomogeneous terms $
k^{(1)}_i $ replaced by 
\bea
\alfapi k^{(1)}_i|_{DIS_\gamma} =  \alfapi k^{(1)}_i - P^{(0)}_{iq} \otimes 
C_\gamma
\ena
as can be immediately derived by expressing the $\MSB$ parton
distributions $F_{i/\gamma} $ in terms of their $DIS_\gamma$
counterparts  
\bea
F_{q/\gamma} &=& F_{q/\gamma}|_{DIS_\gamma} - C_\gamma \nonumber \\
F_{g/\gamma} &=& F_{g/\gamma}|_{DIS_\gamma} 
\label{eq:dis}
\ena
in eq. (\ref{eq:evol}).
Note that the homogeneous terms are not affected by this transformation.
It is shown in \cite{glurevo} that, in this convention, the leading
logarithmic and the beyond leading logarithmic parton distributions
remain very  close to each other over the all range in $x$.

We adopt here a different approach which also absorbs the troublesome
"large" $\ln(1-x)$ terms with the added advantages that the parton
distributions we define are universal (i.e. independent of the reference
process)  and obey the $\MSB$ evolution equations \cite{font}, as all
hadron structure  functions in practical use today.  Furthermore they
are physically motivated by a careful analysis on how to implement the
non-perturbative input. One criticism which may be raised against the
procedure in \cite{glurevo} is the following. At $Q=Q_0$ the structure
function $F^\gamma_2(x,Q_0)$ is entirely given by the non-perturbative
quark and gluon distributions which are identified with the quark and
gluon distributions in a vector meson according to the $VDM$ hypothesis.
But there is no  compelling reason to make this identification in the
$DIS_\gamma$ scheme rather than in the $\MSB$ scheme (the difference is
the term $C_\gamma$) or even to choose another  physical process where
to perform the identification with the $VDM$ input. A careful analysis
\cite{font} shows that at $Q=Q_0$ the relationship between the photon
structure function and the $VDM$ quark distribution should be  in the
$\MSB$ scheme (neglecting higher order terms) 
\bea
F^\gamma_2(x,Q_0) &=& C_\gamma(x) + F^{NP}_{q/\gamma}(x,Q_0) \nonumber \\
&=& C_\gamma(x) -  C_0(x) + F^{VDM}_{q/\gamma}(x,Q_0)
\label{eq:font}
\ena
with 
\bea
C_0(x) = 6\ \alfapi e^4_q \ \left( (x^2+(1-x)^2 \ln(1-x) + 2 x (1-x) \right)
\ena
In words, it means that the $VDM$ hypothesis does not apply to
$F^{NP}_{i/\gamma}$, introduced in eq. (\ref{eq:solu}), but rather to
the combination $F^{VDM}_{q/\gamma} = F^{NP}_{q/\gamma} + C_0$ which is
independent of the regularization scheme. Thus, the dominant part of the
direct term has to be absorbed in the non-perturbative input. The
function $C_0(x)$ arises from an analysis in the collinear approximation
and it reflects the singularity structure associated to the external
photon leg: it is therefore independent of the hard process in which the
photon is involved.  In that sense the definition of our $VDM$ input is
universal, independent of the considered observable. Away from $Q=Q_0$,
the function $F^{NP}_{i/\gamma}$ evolves according to the homogeneous
Altarelli-Parisi equations in the $\MSB$ scheme.

Roughly speaking, it can be said that in the approach of \cite{glurevo}
the direct term is entirely absorbed in the anomalous photon component
with an appropriate modification of the evolution  equation
($DIS_{\gamma}$ scheme) whereas, in our approach \cite{font}, part of
the direct term is absorbed in the non-perturbative input at the scale
$M_0$  and then evolves with it in the usual $\MSB$ scheme.

Throughout this work we use the parton distributions defined in
\cite{font} which are adjusted to reproduce the photon structure
function \cite{jade}-\cite{amy}. They are also in good agreement with
the single jet production cross section in $\gamma \ \gamma$ collisions
at TRISTAN energies \cite{top}, \cite{gaga}. The scale $M_0$ at which
the perturbative input  vanishes is $M^2_0 \simeq m^2_\rho \simeq .5\
GeV^2$. The "standard set" is obtained when the full $VDM$ input is used
in eq. (\ref{eq:font}) but we have the flexibility of changing the
overall normalization of this component so that its effect on physical
quantities can be easily analysed. The $VDM$ input is related to the
pion structure functions of \cite{pion}.

Until now no experimental results have been presented with a fixed
photon-proton energy. All data involve a convolution on the photon
energy. We approximate here the H1 kinematical configuration
\cite{h1jet}  and convolute the  $\gamma \ p$ cross section with the
Weizs\"acker-Williams spectrum  to construct the jet cross section in $e
\ p$ collisions.  Taking into account the  experimental tagging
condition we use 
\bea
F_{\gamma/ e}(z) = {\alpha \over 2 \pi z} \left( 1+(1-z)^2 \right)
   \ln{ (1-z) Q^2_{max} \over z^2\ m^2_e }
\label{eq:antitag}
\ena
where $m_e$ is the electron mass and the maximum virtuality of the
quasi-real photon is given by $Q^2_{max}=.01\ GeV^2$. The photon
momentum is further restricted by the condition $.25 < z < .7$. Let us
just mention that given the smallness of the virtuality of the incoming
photon it is a good approximation to neglect it altogether
\cite{bor,gaga,gaga1}.

For the following studies we use for the proton structure function the
quark and gluon distributions of   ref. \cite{abfow}. We thus have a
consistent set of parton distributions since the pion distributions,
used in the definition of the $VDM$ component, have been derived from an
analysis of $\pi \ p$ data using the parametrization of  \cite{abfow}
for the proton.

\section{Phenomenological studies}

As is well known, all perturbatively calculated cross section suffer
from scale ambiguities. We illustrate this point by plotting for Hera
energy, ${\sqrt {s_{e p}}} = 295 \ GeV$, the variation of the cross
section, at fixed jet $p_T$ ($p_T = 8\ GeV/c$) and pseudo-rapidity
($\eta = 0$). In fig. 1a) the leading logarithmic predictions, $i.e.$
arbitrarily setting $K^D_{i\gamma} = K^{SF}_{i j} = 0$, are shown: the
plot reflects the monotonous variation of the coupling $\alpha_s(\mu)$
when the factorization scale $M$ is fixed and the monotonous variation
of the structure function with $M$ when $\mu$ is fixed. There is no
prefered point on this plot. The picture in the next-to-leading
approximation is quite different (fig. 1b)) and the  2-dimensional
surface now exhibits a saddle point referred to as the "optimal point".
This is seen, more quantitatively, in fig. 1c) which shows the
"equipotential" lines of the surface: in particular one observes a
region of stability extending roughly between $ 2.2 < \mu \ (GeV/c) <
4.5 $ and  $ 2. < M \ (GeV/c) < 12$.  The same pattern holds true at
other values of transverse momentum and pseudo-rapidity relevant for
Hera energies and we find that both $M$ and $\mu$  slightly increase
with the transverse momentum and the rapidity although the precise value
of the scale is not important in the neighborhood of the optimal point.
The features of scale compensation in perturbative calculations have
been extensively studied since the pioneering works of  Stevenson and
Politzer \cite{steve}. In the case of reactions involving real photons
the inhomogeneous term in eq. (\ref{eq:evol}) introduces an added
complication \cite{optim} which has led to some confusion in recent
studies \cite{gorsto}, \cite{krasa}. When calculating the higher  order
corrections to the direct sub-processes ($\gamma \ q \rightarrow g\ q, \
\gamma\ g \rightarrow q\ \bar q$) and requiring one of the final partons
to be at large $p_T$ one has to integrate over a kinematical
configuration where the photon decays,  for exemple,  into a real
anti-quark and a virtual quark which can be on-shell. This leads to a
divergence at the pole of the quark propagator (collinear
configuration). The factorization scale separates the region of high
quark virtuality which contributes to the higher order $K^D_{i\gamma}$
term in eq. (\ref{eq:dir}) and the region of small virtuality which is
associated to the photon structure function and therefore to $\dsigsf$.
Before QCD is turned on ($\alpha_s = 0$) in the evolution equations
(\ref{eq:evol}), the cross section eq. (\ref{eq:sig}) is exactly
independent of the scale $M$ which is just an arbitrary parameter  which
separates the hard regime from the soft (structure function) one. When
QCD is turned on, the $\ln M$ dependence is resummed to all orders in
the structure function while the balancing term in $K^D_{i\gamma}$ is
not,  thereby leading to a residual $M$ dependence in the combination
$\dsigd + \ \dsigsf $ as is expected when working with a truncated
perturbative series: the scale dependence is partially compensated
between the higher order term of eq. (\ref{eq:dir}) and the leading one
in eq. (\ref{eq:sf})  (variation associated to the inhomogeneous term,
$k_i$, in eq. (\ref{eq:evol}))  on the one hand, and between the two
terms on the right hand side of eq.  (\ref{eq:sf}) (homogeneous term in
eq. (\ref{eq:evol})). In no case, it is expected that $\dsigd$ and
$\dsigsf$ be separatly stable under changes  of $M$ and therefore they
cannot be physical quantities to be compared with experiment. An
apparent stability may however  be obtained for $\dsigsf$ when choosing
$M=\mu$, so that the variation of the  coupling partially cancels that
of the photon structure function. 

To further test the practical effect of optimization we show in fig. 2)
the rapidity distribution at fixed $p_T = 8\ GeV/c$ under two
hypotheses: the solid line shows the optimal result while the dashed
line shows the predictions based on the commonly used choice
$M=\mu=p_T$. The variation does not exceeds $8\%$ with a moderate
rapidity dependence: the optimization decreases slightly the predictions
in the backward region ($i.e.$ the photon fragmentation region)  and
incresases them in the forward one. In the following, we keep the
traditional choice $M=\mu=p_T$ which turns out to be quite sufficient
for comparison with present data. Also shown in the figure is the
contribution of the direct term, eq. (\ref{eq:dir}): it is negligible
except in the backward region where it reaches $20\%$ of the full cross
section. Needless to recall  that this result is only qualitative as the
magnitude of the direct term varies much more than the full cross
section under changes of scales.

We show in fig. 3a) the results of our calculation under the H1
experimental conditions. The rapidity distribution is obtained
integrating over $p_T > 7\ Gev/c$ and compared with the data
\cite{h1jet} . The rising rapidity spectrum observed experimentally
cannot be reproduced. Part of the problem  at large $\eta$ may be due to
the contamination of the jet with particles from the "underlying" event
which may play an important role given the rather low jet $p_T$ values
involved. In order to test the sensitivity of our predictions to
structure functions we play the following extreme games: setting the
gluon density in the photon to $0$ gives the dashed curve with a totally
wrong rapidity dependence while setting the gluon in the proton to $0$
essentially decreases the cross section without changing the shape of
the curve in the experimentally accessible domain.  We also show the
predictions assuming a vanishing $VDM$ input: as expected only  the
forward rapidity domain is strongly affected. The only virtue of these
games is to indicate which regions are important to constrain the
various structure  functions and what precision the data should reach to
give quantitative  constraints.

In the same spirit we plot in fig. 3b) the average momentum of the
partons  in the proton, $<x_p>$, and in the electron, $<x_e>$, as
functions of the rapidity, for the cross section shown in fig. 3a). We
note the small value of  $<x_p>$ over the whole range confirming the
important role played by the gluon  in the proton as just seen above.
When Hera data become precise enough they can  effectively be used to
constrain the proton structure function in the vicinity of $x_p \simeq
.05$ at values of $Q^2$ around 50 to 100 $GeV^2$. The average value
$<x_e>$ of the partons in the electron is obtained by convoluting the
photon spectrum eq. (\ref{eq:antitag}) with the parton distribution in
the photon and it is calculated here using only the $SF$ component. The
rather "large" $<x_e>$ obtained is due to the fact that the parton
distributions in  the photon are not as peaked as that in the proton. In
the range considered by H1 the cross section is mostly sensitive to the
quark density which is also probed in studies of $F^\gamma_2$. To
achieve good sensitivity to the gluon distribution would require a
larger rapidity coverage in the forward region.

In fig. 4), we plot the jet distribution in $p_T$ after integration over
the rapidity domain $-1. < \eta < 1.5$. Excellent agreement is reached
both in magnitude and in shape. In view of fig. 3a) however it is clear
that this agreement is purely fortuitous! This illustrates the fact that
rapidity distributions are more powerful than transverse momentum ones
to constrain the  parameters of the theory.

In conclusion, we find the study of jet photoproduction at Hera very
promising. Concerning the quantitative comparison between theory and
experiment it appears that the relatively low values of the jet
transverse momentum do not allow us to exclude the possibility of
contamination of the jets with low $p_T$ fragments of the proton making
the extraction of the parton momentum from the jet momentum very
difficult and the comparison with the theory rather dubious at this
point of the analysis. This shows also the relevance of collecting data
on inclusive particle production \cite{h1part} at large transverse
momentum which do not suffer from such an ambiguity. The extra degrees
of freedom contained in the fragmentation functions are by now rather
well constrained \cite{guil2} and the comparaison with theoretical
calculations \cite{thpart} on particle production would also be very
useful to determine the parton distributions in the photon.

\vs{10 }
{\large{\bf Acknowledgements}}

We thank M. Erdmann for several enlightning discussions. We  are
indebted to the EEC programme "Human Capital and Mobility", Network
"Physics at High Energy Colliders", contract CHRX-CT93-0357 (DG 12 COMA)
for financial support.

\newpage

{\Large{\bf Figure Captions}}

\vspace{9 mm}

\begin{tabular}{ll}
Fig. 1 & Variation of the cross section at fixed $p_T=8\ GeV/c$ and 
	$\eta=0$ \\
 & as a function of the factorization scale $M$ and the renormalization 
	scale $\mu$: \\
 & a) leading logarithmic approximation; b) next-to-leading logarithmic \\
 & approximation; c) contour plot for the NTL approximation. The equal \\
 & cross section lines are labeled in $pb \ (GeV/c)^{-1})$. \\
Fig. 2 & Rapidity dependence of the cross section at fixed $p_T=8 \ GeV/c$: 
 the solid \\
 & curve is the optimized prediction while the dashed curve is
 obtained with \\
 & the choice $M=\mu=p_T$. The dotted curve is the contribution of the
 direct \\
 & term (multiplied by a factor 4). \\
Fig. 3 & a) Rapidity dependence of the theoretical predictions  compared 
to the \\
 & experimental data of H1. The dashed curve is obtained when 
$F_{g/\gamma}=0$ \\
 & and the dotted one when $F_{g/p}=0$. The dash-dotted curve results 
from \\
 & the choice $F^{VDM}_{i/\gamma}=0$ in eq. (\ref{eq:font}). \\
 & b) Average values of the scaled longitudinal momenta in the proton 
and \\
 & the electron. \\
Fig. 4 & Transverse momentum dependence of the cross section integrated 
over \\
 &the pseudo-rapidity range $-1.<\eta<1.5$ and comparison with H1 data. \\

\end{tabular}

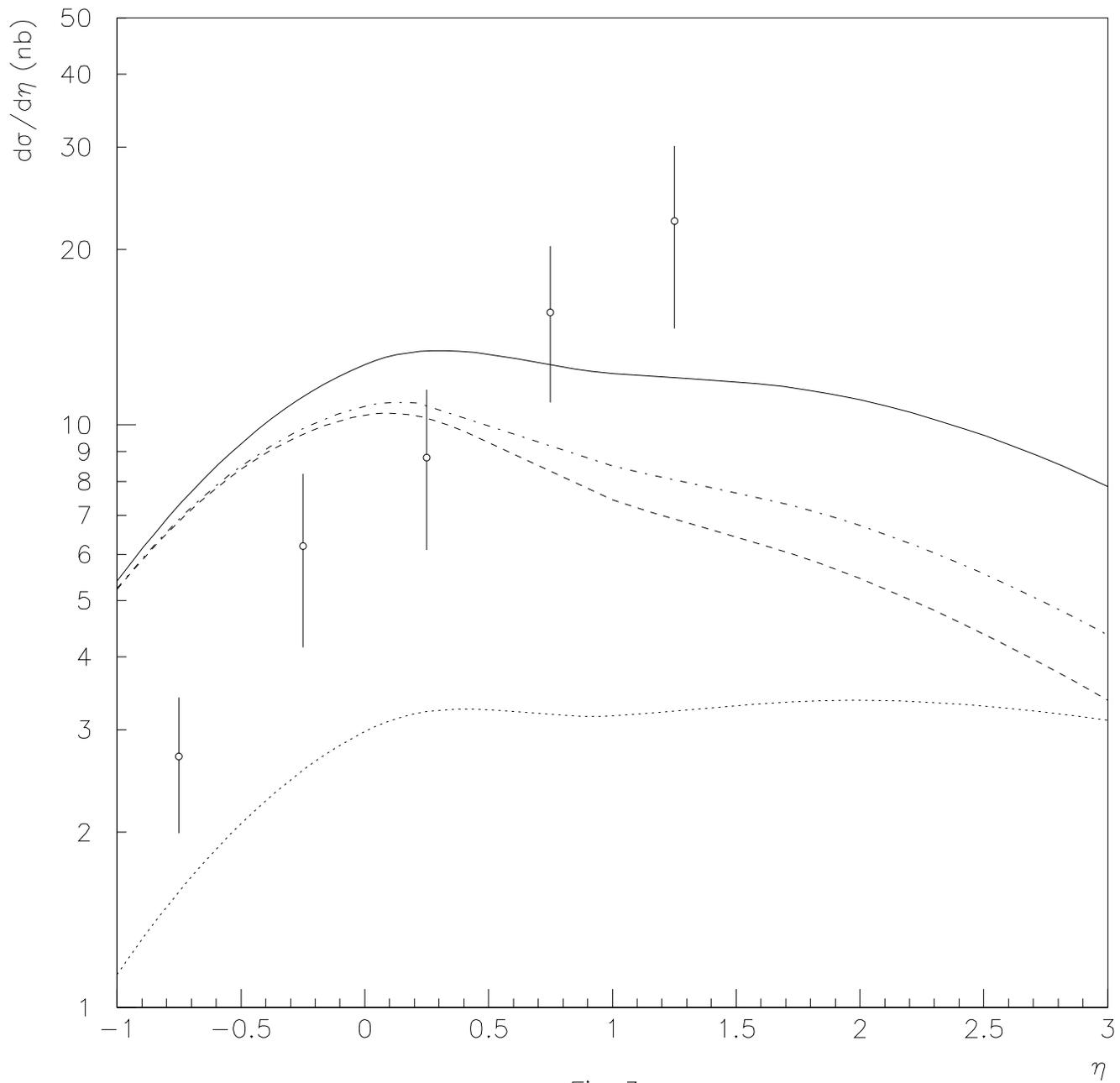
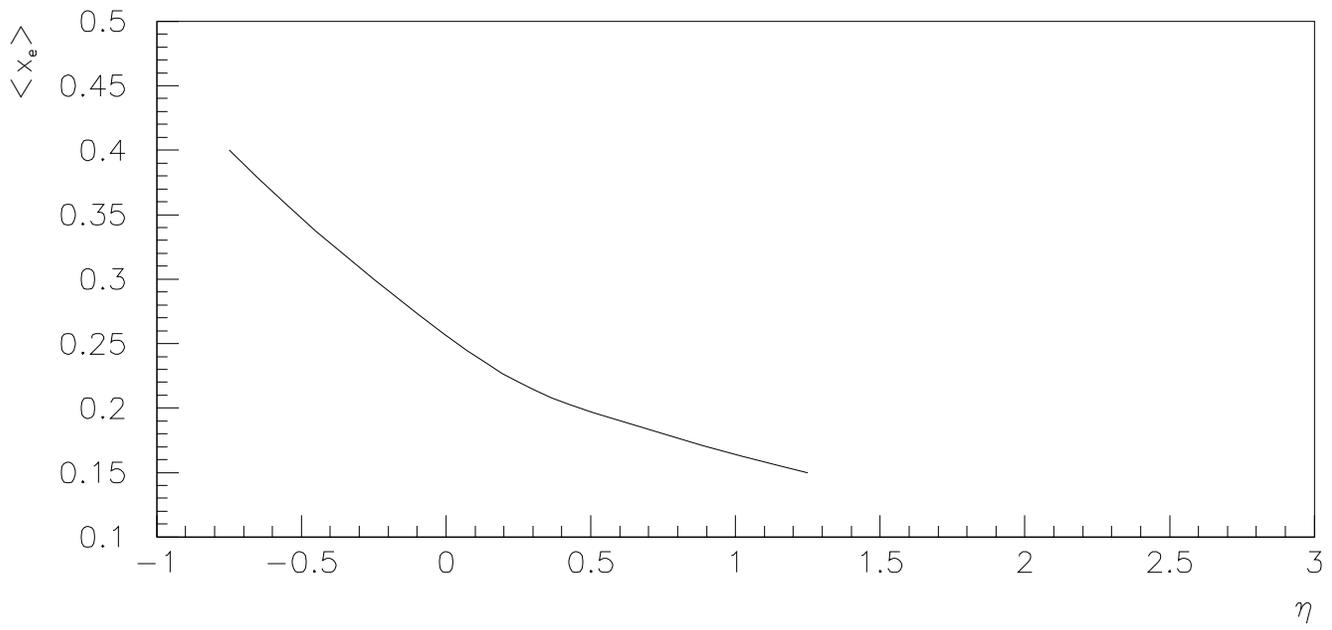
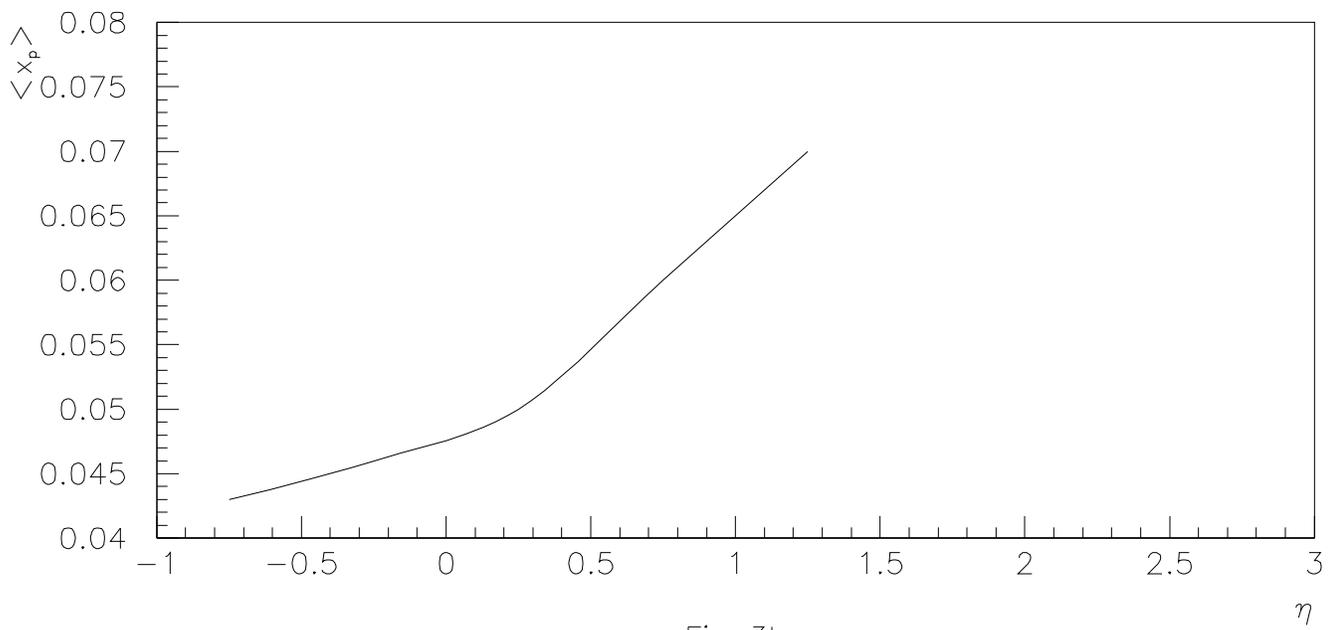
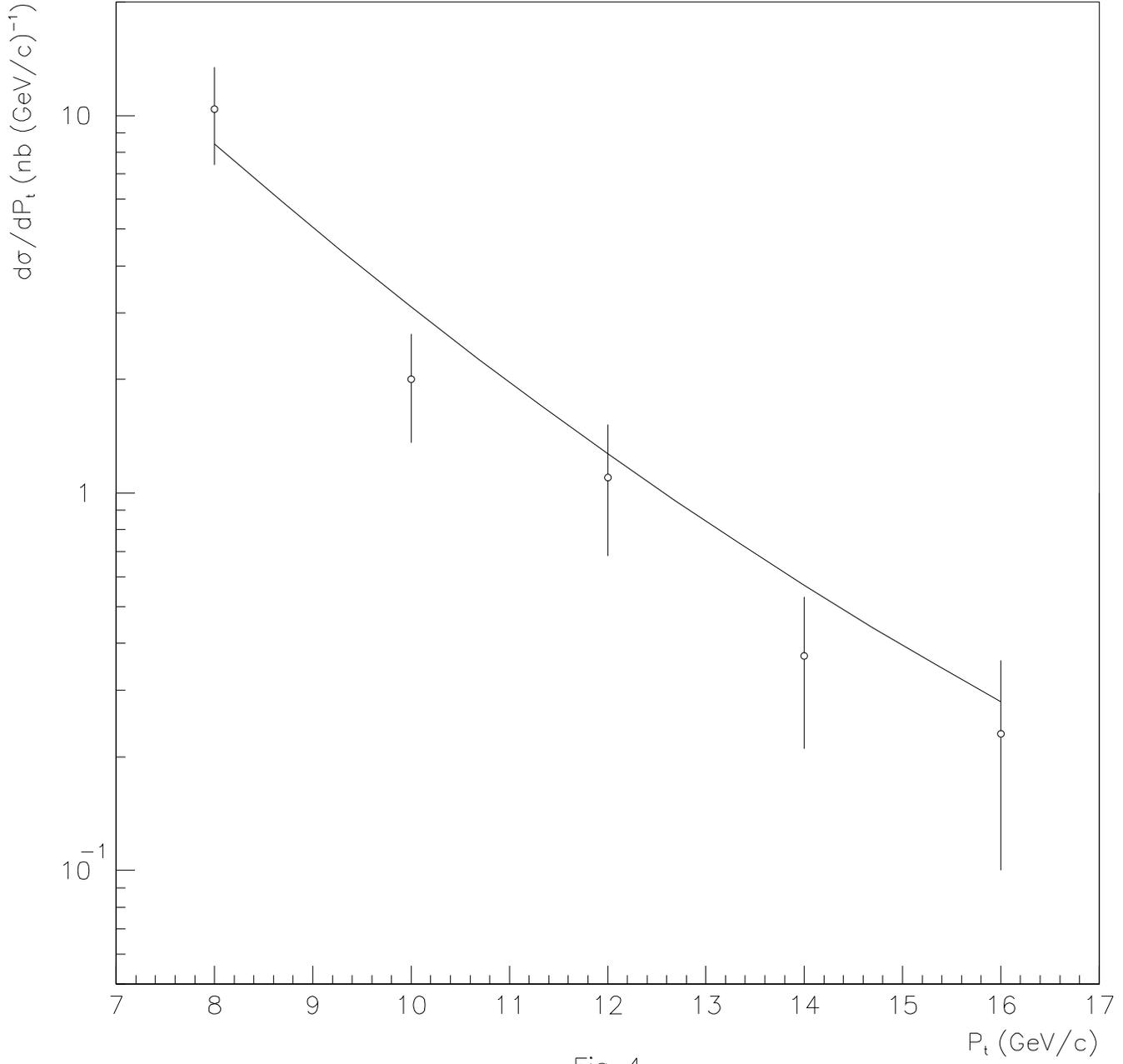
\end{document}